\documentclass{emulateapj}

\usepackage{amsmath}
\usepackage{graphicx}
\usepackage{graphics}

\usepackage{affils}
\usepackage{url}
\usepackage{hyperref}

\DeclareMathSizes{11}{19}{13}{8}

\usepackage{eso-pic}

\AddToShipoutPictureBG*{%
  \AtPageUpperLeft{%
    \hspace{0.65\paperwidth}%
    \raisebox{-3.5\baselineskip}{%
      \makebox[0pt][l]{\textnormal{DES-2016-0181}}
}}}%

\AddToShipoutPictureBG*{%
  \AtPageUpperLeft{%
    \hspace{0.65\paperwidth}%
    \raisebox{-4.25\baselineskip}{%
      \makebox[0pt][l]{\textnormal{FERMILAB-PUB-16-218-AE-PPD}}
}}}%

\slugcomment{In Prep. \today}

\shorttitle{GW151226 Optical Follow-Up}
\shortauthors{Cowperthwaite et~al.}

\begin{document}

\title{A DECam search for an Optical Counterpart to the LIGO Gravitational Wave Event GW151226}

\DeclareAffil{cfa}{Harvard-Smithsonian Center for Astrophysics,
	60 Garden Street, Cambridge, Massachusetts 02138, USA}
\DeclareAffil{gfrp}{NSF GRFP Fellow, e-mail: pcowpert@cfa.harvard.edu}
\DeclareAffil{slac}{SLAC National Accelerator Laboratory, Menlo Park, CA  94025, USA}
\DeclareAffil{fermi}{Fermi National Accelerator Laboratory, P. O. Box 500,
Batavia, IL 60510, USA}
\DeclareAffil{kavli}{Kavli Institute for Cosmological Physics, University of
Chicago, Chicago, IL 60637, USA}
\DeclareAffil{kavlilong}{Enrico Fermi Institute, Department of Physics, Department of 
Astronomy \& Astrophysics, and Kavli Institute for Cosmological Physics, University of
 Chicago, Chicago, IL 60637, USA}
\DeclareAffil{berkeleyTAC}{Department of Astronomy \& Theoretical Astrophysics Center, 
University of California, Berkeley, CA 94720-3411, USA}
\DeclareAffil{berkeleyAstro}{Department of Astronomy, University of California, Berkeley,  501 Campbell Hall, Berkeley, CA 94720, USA}
\DeclareAffil{berkeleyphys}{Departments of Physics and Astronomy, University of California, Berkeley, CA 94720-3411, USA}
\DeclareAffil{LBNL}{Lawrence Berkeley National Laboratory, 1 Cyclotron Road, Berkeley, CA 94720, USA}
\DeclareAffil{steward}{Steward Observatory, University of Arizona, 933 N. 
Cherry Avenue, Tucson, AZ 85721, USA}
\DeclareAffil{columbia}{Columbia Astrophysics Laboratory, Columbia University, Pupin Hall,
New York, NY, 10027, USA}
\DeclareAffil{csic}{Instituto de Fisica Teorica UAM/CSIC,
Universidad Autonoma de Madrid, 28049 Madrid, Spain}
\DeclareAffil{syracuse}{Physics Department, Syracuse University, Syracuse NY 13244, USA}
\DeclareAffil{ohioU}{Astrophysical Institute, Department of Physics and Astronomy, 251B 
Clippinger Lab, Ohio University, Athens, OH 45701, USA}
\DeclareAffil{arizonaNOAO}{National Optical Astronomy Observatory, 
950 North Cherry Avenue, Tucson, AZ, 85719, USA}
\DeclareAffil{illastro}{Astronomy Department, University of Illinois at Urbana-Champaign,
1002 W.\ Green Street, Urbana, IL 61801, USA}
\DeclareAffil{illphys}{Department of Physics, University of Illinois at Urbana-Champaign,
1110 W.\ Green Street, Urbana, IL 61801, USA}
\DeclareAffil{goddard}{Astrophysics Science Division, NASA Goddard Space Flight Center,
8800 Greenbelt Road, Greenbelt, MD 20771, USA}
\DeclareAffil{jsi}{Joint Space-Science Institute, University of Maryland, College Park, MD 20742, USA}
\DeclareAffil{penn}{Department of Physics and Astronomy, University of
Pennsylvania, Philadelphia, PA 19104, USA}
\DeclareAffil{pennlong}{Department of Astronomy \& Astrophysics, Center for Gravitational Wave and Particle Astrophysics, 
and Center for Theoretical and Observational Cosmology, 525 Davey Lab, Pennsylvania State University, University Park, PA 16802, USA}
\DeclareAffil{ICTPbrazil}{ICTP South American Institute for Fundamental Research Instituto 
de F\'{\i}sica Te\'orica, Universidade Estadual Paulista, S\~ao Paulo, Brazil}
\DeclareAffil{IdA}{Laboratorio Interinstitucional de e-Astronomia - LIneA, Rua Gal. Jose Cristino 77, 
Rio de Janeiro, RJ - 20921-400, Brazil }
\DeclareAffil{stsci}{Space Telescope Science Institute, 3700 San Martin Dr., Baltimore, MD 21218, USA}
\DeclareAffil{CTIO}{Cerro Tololo Inter-American Observatory, National Optical Astronomy Observatory, Casilla 603, La Serena, Chile}
\DeclareAffil{ncsa}{National Center for Supercomputing Applications, 1205 West Clark St., Urbana, IL 61801, USA}
\DeclareAffil{UCLondon}{Department of Physics \& Astronomy, University College London, Gower Street, London, WC1E 6BT, UK}
\DeclareAffil{Rhodes}{Department of Physics and Electronics, Rhodes University, PO Box 94, Grahamstown, 6140, South Africa}
\DeclareAffil{princeton}{Department of Astrophysical Sciences, Princeton University, Peyton Hall, Princeton, NJ 08544, USA}
\DeclareAffil{CNRS}{CNRS, UMR 7095, Institut d'Astrophysique de Paris, F-75014, Paris, France}
\DeclareAffil{UPMC}{Sorbonne Universit\'es, UPMC Univ Paris 06, UMR 7095, Institut d'Astrophysique de Paris, F-75014, Paris, France}
\DeclareAffil{kavliStanford}{Kavli Institute for Particle Astrophysics \& Cosmology, P. O. Box 2450, Stanford University, Stanford, CA 94305, USA}
\DeclareAffil{ONBrazil}{Observat\'orio Nacional, Rua Gal. Jos\'e Cristino 77, Rio de Janeiro, RJ - 20921-400, Brazil}
\DeclareAffil{IEEC}{Institut de Ci\`encies de l'Espai, IEEC-CSIC, Campus UAB, Carrer de Can Magrans, s/n,  08193 Bellaterra, Barcelona, Spain}
\DeclareAffil{IFAE}{Institut de F\'{\i}sica d'Altes Energies (IFAE), The Barcelona Institute of Science and Technology, 
Campus UAB, 08193 Bellaterra (Barcelona) Spain}
\DeclareAffil{portsmouth}{Institute of Cosmology \& Gravitation, University of Portsmouth, Portsmouth, PO1 3FX, UK}
\DeclareAffil{southampton}{School of Physics and Astronomy, University of Southampton,  Southampton, SO17 1BJ, UK}
\DeclareAffil{ECU}{Excellence Cluster Universe, Boltzmannstr.\ 2, 85748 Garching, Germany}
\DeclareAffil{ludwig}{Faculty of Physics, Ludwig-Maximilians-Universit\"at, Scheinerstr. 1, 81679 Munich, Germany}
\DeclareAffil{michAstro}{Department of Astronomy, University of Michigan, Ann Arbor, MI 48109, USA}
\DeclareAffil{michPhys}{Department of Physics, University of Michigan, Ann Arbor, MI 48109, USA}
\DeclareAffil{IACambridge}{Institute of Astronomy, University of Cambridge, Madingley Road, Cambridge CB3 0HA, UK}
\DeclareAffil{kavliCambridge}{Kavli Institute for Cosmology, University of Cambridge, Madingley Road, Cambridge CB3 0HA, UK}
\DeclareAffil{OSUAstro}{Center for Cosmology and Astro-Particle Physics, The Ohio State University, Columbus, OH 43210, USA}
\DeclareAffil{OSUPhys}{Department of Physics, The Ohio State University, Columbus, OH 43210, USA}
\DeclareAffil{AAO}{Australian Astronomical Observatory, North Ryde, NSW 2113, Australia}
\DeclareAffil{DFM}{Departamento de F\'{\i}sica Matem\'atica,  Instituto de F\'{\i}sica, Universidade de S\~ao Paulo, 
 CP 66318, CEP 05314-970, S\~ao Paulo, SP,  Brazil}
 \DeclareAffil{texAM}{George P. and Cynthia Woods Mitchell Institute for Fundamental Physics and Astronomy, and 
 Department of Physics and Astronomy, Texas A\&M University, College Station, TX 77843,  USA}
 \DeclareAffil{ICR}{Instituci\'o Catalana de Recerca i Estudis Avan\c{c}ats, E-08010 Barcelona, Spain}
 \DeclareAffil{maxplanck}{Max Planck Institute for Extraterrestrial Physics, Giessenbachstrasse, 85748 Garching, Germany}
 \DeclareAffil{JPL}{Jet Propulsion Laboratory, California Institute of Technology, 4800 Oak Grove Dr., Pasadena, CA 91109, USA}
 \DeclareAffil{sussex}{Department of Physics and Astronomy, Pevensey Building, University of Sussex, Brighton, BN1 9QH, UK}
 \DeclareAffil{CIEMAT}{Centro de Investigaciones Energ\'eticas, Medioambientales y Tecnol\'ogicas (CIEMAT), Madrid, Spain}
 \DeclareAffil{ludwiglong}{Universit\"ats-Sternwarte, Fakult\"at f\"ur Physik, Ludwig-Maximilians Universit\"at M\"unchen, 
 Scheinerstr. 1, 81679 M\"unchen, Germany}
 \DeclareAffil{icecube}{Dept. of Physics and Wisconsin IceCube Particle Astrophysics Center, University of Wisconsin, Madison, WI 53706, USA}
 \DeclareAffil{NYU}{Center for Cosmology and Particle Physics, New York University, 4 Washington Place, New York, NY 10003, USA}
 
\affilauthorlist{
	P.~S.~Cowperthwaite\affils{cfa,gfrp}, 
	E.~Berger\affils{cfa},
	M.~Soares-Santos\affils{fermi},
	J.~Annis\affils{fermi},
	D.~Brout\affils{penn},
	D.~A.~Brown\affils{syracuse},
	E.~Buckley-Geer\affils{fermi},
	S.~B.~Cenko\affils{goddard,jsi},
	H.~Y.~Chen\affils{kavli},
	R.~Chornock\affils{ohioU}
% ...
	\\ (The DES Collaboration)
	}

\begin{abstract} 
  We report the results of a Dark Energy Camera (DECam)
  optical follow-up of the gravitational wave (GW) event GW151226,
  discovered by the Advanced LIGO detectors. Our observations cover 28.8
  deg$^2$ of the localization region in the $i$ and $z$ bands (containing 3\% 
  of the {\tt BAYESTAR} localization probability), starting 10 hours after the 
  event was announced and spanning four epochs at $2-24$ days after the 
  GW detection. We achieve $5\sigma$ point-source limiting 
  magnitudes of $i\approx21.7$ and $z\approx21.5$, with a scatter
  of $0.4$~mag, in our difference images. Given the two day delay, 
  we search this area for a rapidly declining optical counterpart with
  $\gtrsim 3\sigma$ significance steady decline between the first and final
  observations. We recover four sources that pass our selection criteria, 
  of which three are cataloged AGN. The fourth source is offset
  by $5.8$ arcsec from the center of a galaxy at a distance of 187
  Mpc, exhibits a rapid decline by $0.5$ mag over $4$ days,
  and has a red color of $i-z\approx 0.3$ mag. These properties could satisfy
  a set of cuts designed to identify kilonovae. However, this source
  was detected several times, starting $94$ days prior to GW151226,  
  in the Pan-STARRS Survey for Transients (dubbed as PS15cdi) and
  is therefore unrelated to the GW event. Given its long-term 
  behavior, PS15cdi is likely a Type IIP supernova that 
  transitioned out of its plateau phase during our 
  observations, mimicking a kilonova-like behavior. We comment on
  the implications of this detection for contamination in future
  optical follow-up observations. 
\end{abstract}

\keywords{binaries: close -- catalogs -- gravitational waves -- stars:
  neutron -- surveys}

\section{Introduction}
\label{sec:intro}

The Advanced Laser Interferometer Gravitational-Wave Observatory
(LIGO) is designed to detect the final inspiral and
merger of compact object binaries comprised of neutron stars (NS)
and/or stellar-mass black holes (BH) \citep{abb+09}. 
The first LIGO observing run (designated O1)
commenced on 18 September 2015 with the ability to detect binary
neutron star (BNS) mergers to an average distance of $\approx 75$ Mpc, a
forty-fold increase in volume relative to the previous generation of
ground-based GW detectors \citep{LIGO_dist}. On 2015 September 14 LIGO detected 
the first GW event ever observed, GW150914 \citep{abb+16a}.

The waveform of GW150914 was consistent with the inspiral, merger,
and ring-down of a binary black hole (BBH) system ($36+29$ M$_\odot$; 
\citealt{abb+16a}) providing the first observational
evidence that such systems exist and merge. While there are no robust
theoretical predictions for the expected electromagnetic (EM)
counterparts of such a merger, more than 20 teams conducted a wide
range of follow-up observations spanning from radio to $\gamma$-rays, along with neutrino follow up
\citep{abb+16b, adrian+16, annis+16, conn+16, evans+16, kasliwal+16, sav+16, smartt+16, ss+16, tavani+16}.
This effort included deep optical follow-up observations by our group using DECam covering 100 deg$^2$
(corresponding to a contained probability of 38\%~(11\%) of the initial~(final) sky maps)
 -- making this one of the most comprehensive optical 
follow-up campaigns for GW150914 \citep{ss+16,annis+16}. Our search for rapidly
declining transients to limiting magnitudes of $i\approx21.5$ mag
for red $(i-z=1)$ and $i\approx20.1$ mag for blue $(i-z=-1)$ events yielded no counterpart to
GW150914 \citep{ss+16}. One result of the broader multi-wavelength
follow-up campaign is a claimed coincident detection of a weak short
gamma-ray burst (SGRB) from the {\it Fermi}/GBM detector 0.4 s after
the GW event \citep{conn+16}. However, this event was not
detected in INTEGRAL $\gamma$-ray data \citep{sav+16} and was also
disputed in a re-analysis of the GBM data \citep{greiner+16}.

A second high-significance GW event, designated GW151226, was
discovered by LIGO on 2015 December 26 at 03:38:53 UT \citep{abb+16c}. 
This event was also due to the inspiral and merger of a BBH system, consisting 
of $14.2^{+8.3}_{-3.7}$~M$_\odot$ and $7.5^{+2.3}_{-2.3}$~M$_\odot$ black holes at a luminosity 
distance of $d_L = 440^{+180}_{-190}$~Mpc \citep{abb+16c}.
The initial localization was provided as a probability sky map via a private 
GCN circular 38 hours after the GW detection \citep{sing+15}. We initiated
optical follow-up observations with DECam 10 hours later on 2015 December 28,
and imaged a 28.8 deg$^2$ region in the $i$ and $z$ bands during several epochs. Here we report 
the results of this search. In Section~\ref{sec:obs} we discuss the
observations and data analysis procedures. In Section~\ref{sec:analysis} 
we present our search methodology for potential counterparts to GW151226, 
and the results of this search. We summarize our conclusions in Section~\ref{sec:conc}. We perform 
cosmological calculations assuming $H_0 = 67.8$~km  s$^{-1}$ Mpc$^{-1}$,
$\Omega_{\lambda} = 0.69$, and $\Omega_m = 0.31$ \citep{planck15}. 
 Magnitudes are reported in the AB system. 

\section{Observations and Data Reduction} 
\label{sec:obs}
   
GW151226 was detected on 2015 December 26 at 03:38:53 UT by a 
Compact Binary Coalescence (CBC) search pipeline \citep{abb+16c}. The CBC
pipeline operates by matching the strain data against waveform
templates and is sensitive to mergers containing NS and/or BH. 
The initial sky map was generated by the {\tt BAYESTAR} algorithm
and released 38 hours after the GW detection. {\tt BAYESTAR} is a Bayesian 
algorithm that generates a localization sky map based on the parameter
estimation from the CBC pipeline \citep{sing+14,sing+16}. The sky area contained within the
initial 50\% and 90\% contours was 430 deg$^2$ and 1340 deg$^2$,
respectively. A sky map generated by the {\tt LALInference}
algorithm, which computes the localization using Bayesian forward-modeling 
of the signal morphology \citep{veitch+15}, was released on 2016 January 15 UT, 
after our DECam observations had been concluded. The {\tt LALInference} sky map
is slightly narrower than the sky map from {\tt BAYESTAR} with 50\% and 90\% contours 
of 362 deg$^2$ and 1238 deg$^2$, respectively.


        \begin{figure}[t!]
   \centering
   	\vspace{-0pt}
	\includegraphics[width=1.0 \columnwidth]{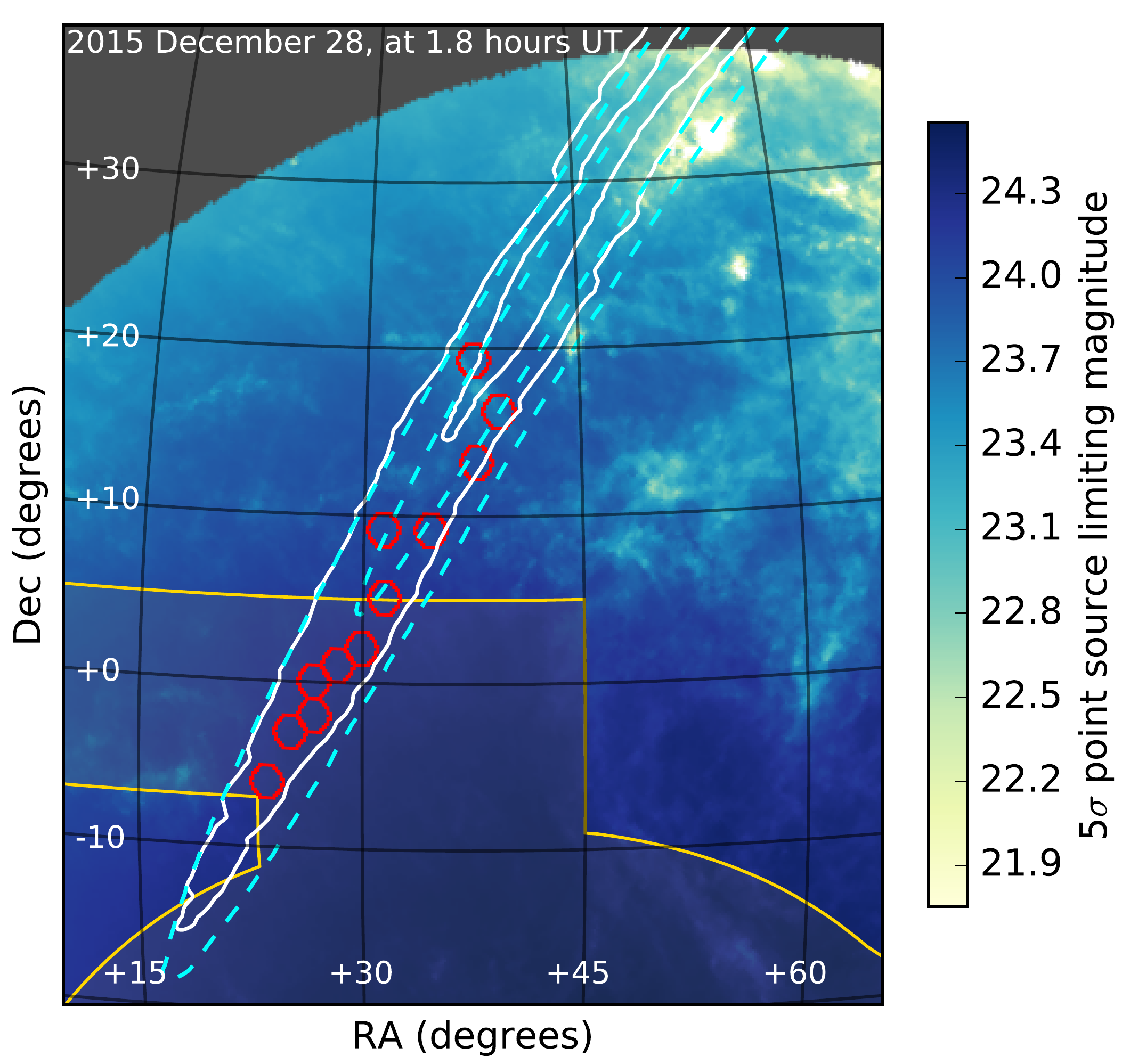}  
	\vspace{-5pt}
        \caption{
        	 Sky region covered by our DECam observations (red
          hexagons) relative to the 50\% and 90\% probability regions 
          from the {\tt BAYESTAR} (cyan contours) and {\tt LALInference}
          (white contours) localization of GW151226. 
          The background color indicates the estimated $5\sigma$ 
          point-source limiting magnitude for a 90~s $i$-band exposure 
          as a function of sky position for the first night of our DECam 
          observations. The variation in the limiting magnitude is largely 
          driven by the dust extinction and airmass at that position. 
          The dark grey regions indicate sky positions that 
          were unobservable due to the telescope pointing limits.  The yellow contour 
          indicates the region of sky covered by the Dark Energy Survey 
          (DES). The total effective area for the 12 DECam pointings is 28.8 
          deg$^2$, corresponding to 3\%~(2\%) of the probability in the
          {\tt BAYESTAR}~({\tt LALInference}) sky map.}
   \label{fig:obs}
   \end{figure}
   
We initiated follow-up observations with DECam on 2015 December 28 UT,
two days after the GW detection and 10 hours after distribution of the {\tt BAYESTAR}
sky map. DECam is a wide-field optical imager with a 3 deg$^2$ field of view 
\citep{flaugher+15}. We imaged a 28.8 deg$^2$ region corresponding to 3\%
of the sky localization probability when convolved with the initial {\tt BAYESTAR} map 
and 2\% of the localization probability in the final {\tt LALInference} sky map. 
The pointings and ordering of the DECam observations were determined using
the automated algorithm described in Soares-Santos et al. \citeyearpar{ss+16}. The choice 
of observing fields was constrained by weather, instrument availability, and the available
time to observe this sky region given its high airmass. We
obtained four epochs of data with each epoch consisting of one 90 s
exposure in $i$-band and two 90 s exposures in $z$-band for each of the 12 pointings. The first
epoch was obtained 2--3 days after the GW event time (2015
December 28--29 UT), the second epoch was at 6 days (2016 January 1
UT), the third epoch was at 13--14 days (2016 January 8--9), and the
fourth epoch was at 23--24 days (2016 January 18--19). A summary of
the observations is provided in Table~\ref{datatable} and a visual
representation of the sky region is shown in
Figure~\ref{fig:obs}.
  
We processed the data using an implementation of the {\tt photpipe} 
pipeline modified for DECam images. {\tt Photpipe} is a
pipeline used in several time-domain surveys (e.g.,
SuperMACHO, ESSENCE, Pan-STARRS1; see \citealt{rest+05,garg+07,mik+07,rest+14}), designed to perform
single-epoch image processing including image calibration (e.g., bias
subtraction, cross-talk corrections, flat-fielding), astrometric
calibration, image coaddition, and photometric calibration.
Additionally, {\tt photpipe} performs difference imaging using {\tt
  hotpants} \citep{alard00,becker15} to compute a spatially varying convolution kernel,
followed by photometry on the difference images using an
implementation of {\tt DoPhot} optimized for point spread function (PSF) photometry on
difference images \citep{schechter+93}. Lastly, we
use {\tt photpipe} to perform initial candidate searches by
specifying a required number of spatially coincident detections over a
range of time. Once candidates are identified, {\tt photpipe} performs
``forced" PSF photometry on the subtracted images at the fixed coordinates of
an identified candidate in each available epoch.

In the case of the GW151226 observations, we began with raw images
acquired from the NOAO archive\footnote{http://archive.noao.edu/} and the
most recent calibration files\footnote{http://www.ctio.noao.edu/noao/content/decam-calibration-files}.
Astrometric calibration was performed relative to the Pan-STARRS1 (PS1)
$3\pi$ survey and 2MASS $J$-band catalogs. The two $z$-band exposures
were then coadded. Photometric calibration was performed using the
PS1 $3\pi$ survey with appropriate calibrations between 
PS1 and DECam magnitudes \citep{scolnic+15}. Image subtraction was performed
using observations from the final epoch as templates. The approach
to candidate selection is described in Section~\ref{sec:analysis}. 

\begin{deluxetable*}{lrccccccc}
\tabletypesize{\footnotesize}
\tablecolumns{9}
\tablewidth{0pt}
\tablecaption{Summary of DECam Observations of GW151226
\label{datatable}}
\tablehead{	
    \colhead{Visit}    &  
    \colhead{UT}       & 
    \colhead{$\Delta t$\tablenotemark{a}} &
    \colhead{$\langle$PSF$_i$$\rangle$}    &
    \colhead{$\langle$PSF$_z$$\rangle$}    &
    \colhead{$\langle$airmass$\rangle$}  &
    \colhead{$\langle$depth$_i\rangle$}   &
    \colhead{$\langle$depth$_z\rangle$}  &
    \colhead{$A_{\mathrm{eff}}$\tablenotemark{b}} \\
    \colhead{ } & 
    \colhead{ } &
    \colhead{(days)} &
    \colhead{(arcsec)} &
    \colhead{(arcsec)} &
    \colhead{ } &
    \colhead{(mag)} &
    \colhead{(mag)} &
    \colhead{(deg$^2$)} 
}
\startdata
 Epoch 1 & 2015-12-28.11 & 1.96 & 0.97 & 0.99 & 1.95 & 22.39 & 22.23 & 14.4 \\      
               & 2015-12-29.11 & 2.96 & 1.00 & 0.97 & 1.78 & 22.57 & 22.46 & 14.4 \\      
                \hline
 Epoch 2 & 2016-01-01.06 & 5.91 & 0.95 & 0.90 & 1.57 & 21.37 & 21.06 & 28.8 \\      
 \hline
 Epoch 3 & 2016-01-08.11 & 12.96 & 1.68 & 1.62 & 2.15 & 22.09 & 21.70 & 24.0 \\      
                & 2016-01-09.11 & 13.96 & 1.17 & 1.12 & 1.80 & 22.44 & 22.17 &  4.8 \\      
                 \hline
 Epoch 4 & 2016-01-18.03 & 22.88 & 1.21 & 1.20 & 1.48 & 22.00 & 22.01 & 12.0 \\      
               & 2016-01-19.01 & 23.86 & 1.29 & 1.25 & 1.71 & 21.86 & 21.90 & 16.8 \\ 
\enddata
\tablecomments{Summary of our DECam follow-up observations of
  GW151226. The PSF, airmass, and depth are the average values across
  all observations on that date. The reported depth corresponds to the mean 
  $5\sigma$ point source detection in the coadded search images.}
\tablenotetext{a}{Time elapsed between the GW trigger time and the time of the first image.} 
  \tablenotetext{b}{The effective area corresponds to 12 DECam pointings 
  taking into account that $\approx 20\%$ of the 3
  deg$^2$ field of view of DECam is lost due to chip gaps (10\%), 3
  dead CCDs (5\%, \citealt{diehl+14}), and masked edge pixels (5\%).}
\end{deluxetable*}

Our observations achieved average $5\sigma$ point-source limiting
magnitudes of $i\approx22.2$ and $z\approx21.9$ in the coadded single-epoch
search images, and $i\approx21.7$ and $z\approx21.5$ in the difference
images, with an epoch-to-epoch scatter of 0.4 mag. The variability in 
depth is driven by the high airmass and
changes in observing conditions, particularly during the second epoch.
 
\section{Search for an Optical Counterpart}
\label{sec:analysis}

The primary focus of our search is a fast-fading transient. 
While the merger of a BBH system is not expected
to produce an EM counterpart, it is informative to consider the
possibility of optical emission due to the presence of some matter in
the system. As a generic example, we consider the behavior of a
transient such as a short gamma-ray burst (SGRB) with a typical beaming-corrected energy of
$E_j \approx 10^{49}$ erg and an opening angle of $\theta_j\approx
10^\circ$ \citep{berger14,fong+15}. If viewed far off-axis $(\theta_{\rm
  obs}\gtrsim 4\theta_j)$ the optical emission will reach peak
brightness after several days, but at the distance of GW151226
($\approx440$ Mpc, \citealt{abb+16c}), the peak brightness will be $i\approx 26$ mag (see Figure 5 of
\citealt{metzger12}), well beyond our detection limit. If the source
is observed moderately off-axis or on-axis $(\theta_{\rm obs} \lesssim
2\theta_j)$, then the light curve will decline throughout our
observations, roughly as $F_\nu\propto t^{-1}$, and will be detectable
at $i\approx$ 21--22 mag in our first observation (see Figures 3 and 4
of \citealt{metzger12}). We can apply a similar argument to the
behavior of a more isotropic (and non-relativistic) outflow given that
any material ejected in a BBH merger is likely to have a low mass and
the outflow will thus become optically thin early, leading to fading
optical emission. Based on this line of reasoning, we search our data
for steadily declining transients.

We identify relevant candidates in the data using the following
selection criteria with the forced photometry from {\tt
  photpipe}. Unless otherwise noted these criteria are applied to
  the $i$-band data due to the greater depth in those observations.

\begin{enumerate}

\item We require non-negative or consistent with zero 
(i.e., within 2$\sigma$ of zero) 
$i$- and $z$-band fluxes in the difference photometry
across all epochs to eliminate any sources that re-brighten in the 
fourth (template) epoch. This provides an initial sample of 602 candidates.

\item We require $\ge 5\sigma$ $i$- and $z$-band detections in the
first epoch and at least one additional $\ge 5\sigma$ $i$-band
detection in either of the two remaining epochs (to eliminate contamination
from asteroids). This criterion leaves a sample of 98 objects.

\item We require a $\ge 3\sigma$ decline in flux between the first and
third epochs to search for significant fading\footnote{We note that this 
criterion effectively requires the detection in the first epoch to be $\gtrsim 5\sigma$ 
producing an effectively shallower transient search. Soares-Santos et al. 
\citeyearpar{ss+16} quantified this effect by injecting fake sources into 
their observations to determine the recovery efficiency and loss of detection
depth from analysis cuts. Here, we forego 
such analysis to focus the discussion on the effects of contamination in 
optical follow-up of GW events.}. We calculate $\sigma$ as the 
quadrature sum of the flux errors from the first and third epochs 
($\sigma = \sqrt{\sigma_1^2 + \sigma_3^2}$, where $\sigma_1$ and
$\sigma_3$ are the flux errors from the first and third epochs,
respectively). This criterion leaves a sample of 48 objects.

\item We reject sources that exhibit a significant ($\ge 3\sigma$) rise in flux 
between the first and second epochs or the second and third epochs to eliminates variable 
sources that do not decline steadily. This criterion leaves a sample of 32 objects. 


\item The remaining 32 candidates from step 4 undergo visual inspection. We reject sources
that are present as a point source in the fourth (template) epoch that do not have a galaxy
within 20\arcsec. Sources are cross-checked against NED\footnote{\url{https://ned.ipac.caltech.edu/}} 
and SIMBAD\footnote{\url{http://simbad.u-strasbg.fr/simbad/}}. This criterion is designed to remove
variable stars and long-timescale transients.
\end{enumerate}

Only four events passed our final criterion. We find that two of those events are 
coincident with the nuclei of known AGN (PKS\,0129-066 and Mrk\,584), indicating 
that they represent AGN variability. A third candidate is coincident with the nucleus of
the bright radio source PMN\,J0203+0956 ($F_\nu(365\,{\rm MHz})\approx 0.4$ Jy, 
\citealt{douglas+96}), also suggesting AGN variability.

\begin{figure*}[t!]
   \centering
	\includegraphics[width=0.95\textwidth]{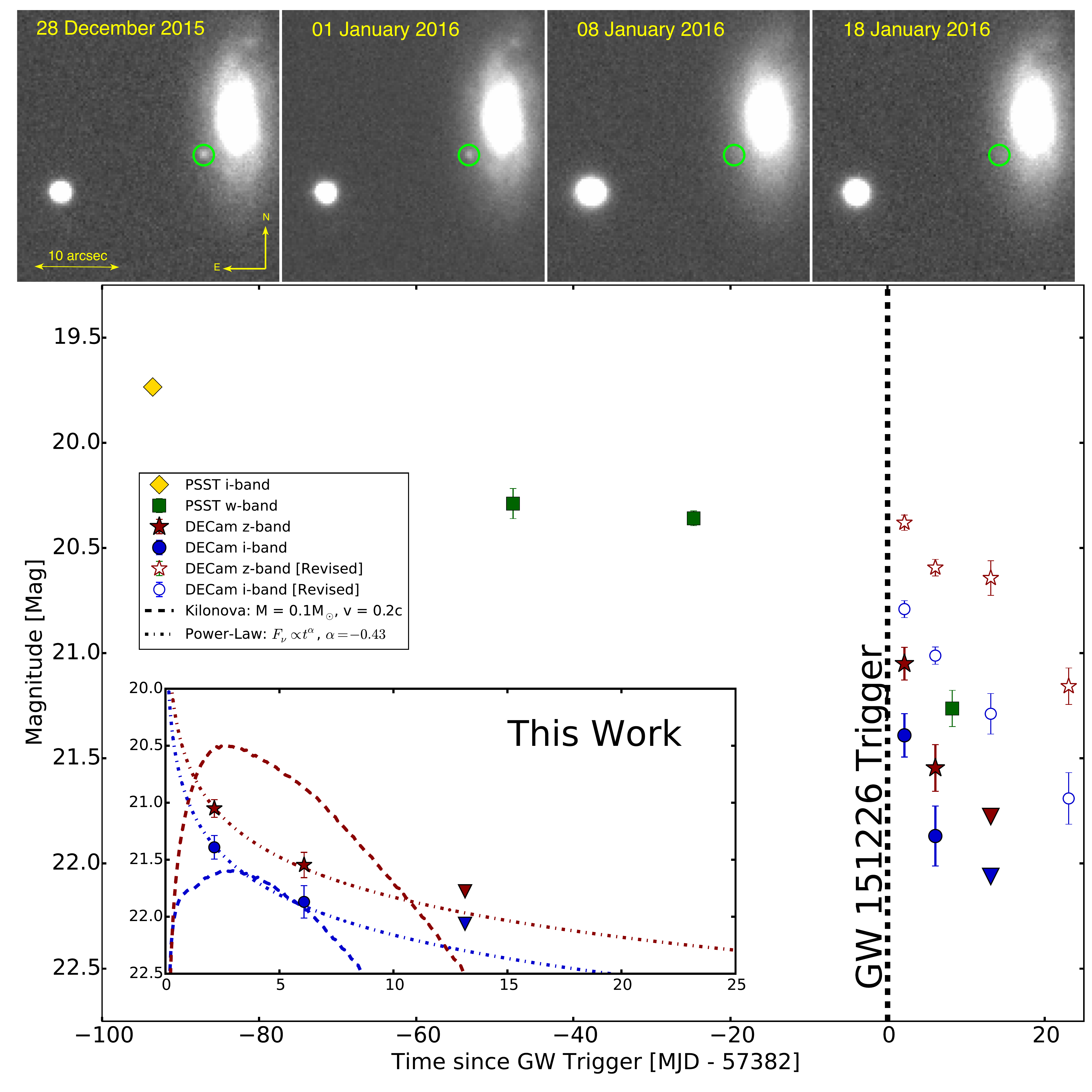}
        \caption{{\it Top}: Single-epoch images of our main candidate
          from all four epochs (green circle). This is the event discovered as PS15cdi in
          the PSST about 94 d prior to GW151226. {\it Bottom}: Light
          curve data for PS15cdi from PSST $w$- and $i$-band
          observations (green squares and yellow diamonds, respectively).
          Our DECam $i$- and $z$-band data are shown as
          blue circles and red stars, respectively. The revised DECam analysis 
          using pre-existing templates is shown as open symbols. Upper
          limits are indicated by triangles. The inset focuses on our
          DECam data, indicating a rapid decline in both $i$ and $z$
          bands.  We fit a power-law model to the data finding a temporal index of 
          $\alpha = -0.43$ (dashed-dot line). Kilonova models from Barnes and 
          Kasen \citeyearpar{barnes13} with $v_{\rm ej}=0.2c$ and $M_{\rm ej}=0.1$ M$_\odot$ 
          at a distance of $187$~Mpc are also shown (dashed line).}
   \label{fig:PS15cdi}
\end{figure*}

The final candidate in our search is located at RA = 01$^{\rm h}$42$^{\rm m}$16$^{\rm s}$.17 
and DEC = $-$02\arcdeg13\arcmin42 6\arcsec~(J2000), with an offset of $5.8$ arcsec from the galaxy 
CGCG 386-030 (RA =  01$^{\rm h}$42$^{\rm m}$15$^{\rm s}$.6, DEC = $-$02\arcdeg13\arcmin38 
5\arcsec; J2000), at $z = 0.041$ or $d_L \approx 187$ Mpc (6dFGS, \citealt{jones+04,jones+09}); see 
Figure~\ref{fig:PS15cdi}. We note that this distance is inconsistent with the 90\% confidence interval for
the distance to GW151226 based on the GW data \citep{abb+16c}. We observe this source in a state of rapid decline with an 
absolute magnitude of $M_i \approx -15$~mag on 2015 December 28 and $M_i\approx -14.5$~mag on 2016 
January 1, indicating a decline rate of $\approx 0.12$ mag d$^{-1}$; the decline rate in $z$-band is 
$\approx 0.10$ mag d$^{-1}$. Additionally, the source exhibits a red $i-z$ color of $0.3$ mag. We 
fit these data to a power-law model typical for GRB afterglows $(F_\nu \propto \nu^{\beta} t^{\alpha})$ 
and find a temporal index of $\alpha = -0.43\pm0.12$ and a spectral index of $\beta = -1.8\pm0.8$, 
both of which differ from the expected values for GRB afterglows ($\alpha \approx -1$, 
$\beta \approx -0.75$, \citealt{sari+98}). Additionally, we compare our observations to a 
kilonova model with ejecta parameters of $v_{\rm ej} = 0.2$c and $M_{\rm ej} = 0.1$ M$_\odot$
 \citep{barnes13}. We find that the timescale of the transient agrees with those expected for kilonovae, 
 but the color is bluer than the expected value of $i-z \approx 1$~mag \citep{barnes13}. Thus, the properties
 of this transient differ from those of GRB afterglows or kilonovae. The observations 
 and models are shown in Figure~\ref{fig:PS15cdi}.

This source was previously detected as PS15cdi on 2015 September 23 by the Pan-STARRS 
Survey for Transients (PSST\footnote{\url{http://psweb.mp.qub.ac.uk/ps1threepi/psdb/candidate/1014216170021342600/}},
\citealt{huber+15}); see Figure~\ref{fig:PS15cdi}. The absolute $i$-band 
magnitude in the first PSST epoch, $M_i\approx -16.6$ mag and the shallow decline of $\approx0.6$~mag 
over $\approx70$~d, are consistent with a Type IIP core-collapse supernova (SN). 
A likely interpretation of the rapid decline in our observations is that PS15cdi is a Type IIP SN undergoing 
the rapid transition from the hydrogen recombination driven plateau to the radioactive $^{56}$Co dominated 
phase \citep{kasen09,dhun+16,sanders+15}. The red $i-z$ color in our data is consistent with observations 
of other IIP SN during this phase of evolution (e.g., SN2013ej, \citealt{dhun+16}).
This transition typically occurs about 100~d post explosion \citep{kasen09,dhun+16,sanders+15}, consistent with the 
timing of our observations relative to the first detection in PSST.

To mitigate the effect of excess flux from PS15cdi still present in our template observations, we repeat the analysis 
using as templates archival DES $i$- and $z$-band images from 2013 December 19. These data were processed 
and image subtraction was performed as described in Section~\ref{sec:obs}. We find that flux from PS15cdi is indeed still present 
in our original template image, leading to revised first epoch absolute magnitudes of $M_i \approx -15.6$ and 
$M_z \approx -16$~mag, and a decline rate between the first and fourth epochs of 0.04~mag d$^{-1}$, in both 
$i$- and $z$-bands. The transient still exhibits a red $i-z$ colors of $\approx 0.4$~mag across all four epochs. 

 
Clearly, we can rule out this candidate based on the PSST detections prior to 
GW151226, but without this crucial information this candidate
would have been a credible optical counterpart based on its light curve behavior and distance. 
It is therefore useful to develop an understanding of the expected rates for such contaminants
to inform expectations in future searches. We adopt a local core-collapse SN rate of  
$7\times10^{-5}$~yr$^{-1}$~Mpc$^{-3}$ \citep{li+11,capp+15}, 
and a Type IIP SN fraction of 48\% of this rate \citep{smith+11}. The rapid
decline phase typically lasts about $20$~d \citep{kasen09,dhun+16,sanders+15}, 
so we consider events that occur within that time frame. Lastly, given its apparent brightness, 
we assume that PS15cdi represents the approximate maximum distance to which we can observe
these events in our data. We thus find an expected occurrence rate of $\sim 0.04$ events in our search area
making our detection of PS15cdi somewhat unlikely, and indicating that $\lesssim 1$ such 
events are expected in a typical GW localization region.

Our detection of PS15cdi clearly demonstrates the presence and impact
of contaminants when conducting optical follow-up of GW events.
Core-collapse SNe are generally not considered to be a
significant contaminant due to their much longer timescales compared
to kilonovae (e.g., \citealt{cowp15}). However, a
source like PS15cdi, caught in a rapid phase of its evolution 
despite its overall long timescale, and
exhibiting a relatively red color could
satisfy a set of criteria designed for finding kilonovae ($\Delta m\gtrsim
0.1$ mag d$^{-1}$ and $i-z\gtrsim 0.3$ mag; \citealt{cowp15}). 

The most effective approach to deal with contaminants like PS15cdi is rapid,
real-time identification. Once a candidate is deemed interesting, optical spectroscopy and 
NIR photometry can quickly distinguish between a SN or kilonova/afterglow
origin. Specifically, the kilonova spectrum will be redder, with clear suppression below $\sim 6000$~\AA~due
to the opacities of $r$-process elements \citep{kasen+13}. By comparison, the SN spectrum will appear bluer and
dominated by iron group opacities \citep{kasen+13}, while the afterglow spectrum will exhibit a featureless power-law
spectrum \citep{berger14}. If pre-existing templates are not available then the significant aspect is rapid initiation of
follow-up observations at $\lesssim 1$ d that can distinguish the rising phase of a kilonova or 
off-axis GRB from a declining SN.

\section{Conclusions}
\label{sec:conc}
We presented the results of our deep optical follow-up of 
GW151226 using the DECam wide-field
imager. Our observations cover a sky area of 28.8 deg$^2$,
corresponding to $3\%$ of the initial {\tt BAYESTAR} probability map
and 2\% of the final {\tt LALInference} map. We obtained
four epochs of observations starting 10 hours after the 
event was announced and spanning 2--24 days post
trigger, with an average $5\sigma$ point-source sensitivity of 
$i\approx21.7$ and $z\approx21.5$, with an epoch-to-epoch scatter
of 0.4 mag, in our difference images.
 
Using the final epoch as a template image, we searched for sources
that display a significant and steady decline in brightness
throughout our observations, and which are not present in the template
epoch. This search yielded four transients, of which three result from
 AGN variability. The final event is located at a distance of about
187 Mpc offset by $5.8''$ from its host galaxy. It also broadly possesses the
observational features of a kilonova in terms of its rapid decline and
red $i-z$ color. However, this source corresponds to the transient
PS15cdi, which was discovered in PSST about 94 days prior to the GW
trigger. It is a likely Type IIP supernova, which our observations
caught in the steep transition at the end of the plateau phase. The
detection of this event indicates that careful rejection of
contaminants, preferably in real time, is essential in order to avoid
mis-identifications of optical counterparts to GW sources.

{\acknowledgements 
P.S.C. is grateful for support provided by the NSF
through the Graduate Research Fellowship Program, grant DGE1144152.
R.J.F.\ gratefully acknowledges support from NSF grant AST--1518052 and 
the Alfred P.\ Sloan Foundation. D.E.H.\ was supported by NSF CAREER grant 
PHY-1151836. He also acknowledges support from the Kavli Institute for  
Cosmological Physics at the University of Chicago through NSF grant  
PHY-1125897 as well as an endowment from the Kavli Foundation.

This research uses services or data provided by the NOAO Science 
Archive. NOAO is operated by the Association of Universities for
Research in Astronomy (AURA), Inc. under a cooperative agreement with
the National Science Foundation. The computations in this paper were
run on the Odyssey cluster supported by the FAS Division of Science,
Research Computing Group at Harvard University. This research has made
use of the NASA/IPAC Extragalactic Database (NED) which is operated by
the Jet Propulsion Laboratory, California Institute of Technology,
under contract with the National Aeronautics and Space Administration.
Light curve data for PS15cdi were obtained from The Open Supernova
Catalog \citep{openSN}. Some of the results in this paper have been 
derived using the HEALPix package \citep{gorski+05}.

Funding for the DES Projects has been provided by the DOE and NSF (USA), 
MEC/MICINN/ MINECO (Spain), STFC (UK), HEFCE (UK). NCSA (UIUC), 
KICP (U. Chicago), CCAPP (Ohio State), MIFPA (Texas A\&M), CNPQ, FAPERJ, 
FINEP (Brazil), DFG (Germany) and the Collaborating Institutions in the Dark Energy Survey.

The Collaborating Institutions are Argonne Lab, UC Santa Cruz, 
University of Cambridge, CIEMAT-Madrid, University of Chicago, University College London, 
DES-Brazil Consortium, University of Edinburgh, ETH Z{\"u}rich, Fermilab, 
University of Illinois, ICE (IEEC-CSIC), IFAE Barcelona, Lawrence Berkeley Lab, 
LMU M{\"u}nchen and the associated Excellence Cluster Universe, 
University of Michigan, NOAO, University of Nottingham, Ohio State University, University of 
Pennsylvania, University of Portsmouth, SLAC National Lab, Stanford University, 
University of Sussex, Texas A\&M University, and the OzDES Membership Consortium.

The DES Data Management System is supported by the NSF under 
Grant Number AST-1138766. The DES participants from Spanish institutions are partially 
supported by MINECO under grants AYA2012-39559, ESP2013-48274, FPA2013-47986, 
and Centro de Excelencia Severo Ochoa SEV-2012-0234. Research leading 
to these results has received funding from the ERC under the EU's 7$^{\rm th}$ 
Framework Programme including grants ERC 240672, 291329 and 306478.
}

\end{document}